\documentclass[preprint, prX]{revtex4}

\usepackage{amsmath}    % need for subequations
\usepackage{graphicx}   % need for figures
\usepackage{verbatim}   % useful for program listings
\usepackage{color}      % use if color is used in text
\usepackage{subfigure}  % use for side-by-side figures
\usepackage{hyperref}   % use for hypertext links, including those to external documents and URLs
\usepackage{amssymb}    % Necessary for \Bbb{R}, etc.
\usepackage{epsfig}
\usepackage{graphics,graphicx}
\usepackage{setspace}
\usepackage{url}
\usepackage{algorithm,algorithmic}

\begin{document}

\begin{abstract}
One of the biggest challenges for the Computer Science of today can be summed up by the paradigm "access to information from $everywhere$ at $anytime$". This is especially true for pervasive computing. With the growth of mobile devices (e.g., smart-phones), on the one hand, and the quick development of the Internet (this has become the really pervasive network of today), on the other hand, the development of real-time pervasive applications has broadened. This paper focuses on the problem of drafting detection in the Ironman triathlons which causes serious problems for the majority of organizers regarding such competitions. A concept of drafting detection system in Ironman is based on the paradigm of pervasive computing. Results of performing a test system show that this concept can along with further development of computer technologies become a reality in the near future.

\textit{To cite paper as follows: I. Fister, I.Fister Jr., Concept of drafting
detection system in Ironmans, Electrotechnical Review, vol. 78, no. 4, pp. 217-222, 2011.
}

\end{abstract}

\title{Concept of drafting detection system in Ironmans}

\author{Iztok Fister}
\altaffiliation{University of Maribor, Faculty of electrical engineering and computer science
Smetanova 17, 2000 Maribor}
\email{iztok.fister@uni-mb.si}

\author{Iztok Fister Jr.}
\altaffiliation{University of Maribor, Faculty of electrical engineering and computer science
Smetanova 17, 2000 Maribor}
\email{iztok.fister@guest.arnes.si}

\maketitle

\section{Introduction}
\indent Pervasive computing (also known as ubiquitous computing) emerged in 1991 with the Weiser's~\cite{weiser:1991} ideas about the computer of the 21st century. Consistent with this idea, people were placed at the fire-front of events and the computer tended to be pushed into the background. In this sense, a notion of disappearing hardware was put into effect. Virtual computers still gave users an opportunity to interact with information devices but they appeared small and specialized. However, this appearance was contradictory with classical, big and universal personal computers.

Pervasive computing has only become a reality quite recently, mostly because of the rapid development of wireless networks and mobile devices. These mobile devices use wireless infrastructure as an access point to information services on the Internet. This information has become accessible from $everywhere$. The Internet assures continued activity. Therefore, information can be accessed from $anywhere$ as well. However, the results of these services are dependent on a $context$, i.e. who, where, when, and why someone has requested information at a defined moment.

Today smart-phones have become the most important mobile devices for accessing the Internet because their feasibility of being connected to wireless networks together with an efficient processing unit. Besides universal phone calling and accessing information services, they also enable additional features such as the global positioning system (GPS), accelerometer, compass, etc. This paper focuses on the global positioning feature used in development of a concept of drafting detection system in Ironmans.

Ironman is a long distance triathlon consisting of three marathons~\cite{petschnig:2007}:
\begin{itemize}
  \item swimming (3.8 kilometers),
  \item cycling (180 kilometers) and
  \item running (42.2 kilometers).
\end{itemize}

On these marathon courses, competitors compete as individuals, i.e. without any help from other competitors. Some competitors, however, often violate this fair-play rule because they wish to improve their performances. This violation is prevalent during cycling, where competitors ride their bicycles within close-knit groups, thus achieving a higher speed and saving energy for later efforts. This phenomenon is known also under the name $drafting$ (or $slipstreaming$). In such group of competitors, the hardest work is performed by the leading competitor who sets the pace, whilst the others purposely shadow.

This kind of bicycle riding in groups is punishable by a 5 minute penalty for the drafting competitor by the World Triathlon Association (WTC) rules. However, detection of drafting is difficult. Today, referees on motorcycles are responsible for drafting detection. However, they can cover only a particular part of the course at a time and estimate violations very subjectively. Therefore, an automatic solution is necessary to be help for referees to detect drafting.

This paper introduces a concept of drafting detection system in Ironmans. The concept consists of two parts:
\begin{itemize}
  \item a pervasive application running on a mobile device and
  \item context-aware web service running on an Internet server.
\end{itemize}

In this concept, the mobile application acts as a gateway that by means of the GPS feature obtains information about the current position of a competitor's bicycle and transmits it to the web service. By knowing positions of other competitors, the web service can calculate whether a competitor is drafting, and if so, how long the violation has been taking place. Knowing the duration of this violation enables separating the drafting violation from normal overtaking. In fact, a drafting violation occurs if overtaking is longer than 20 seconds.

In order to prove the concept of drafting detection system in Ironman a test system was developed. The obtained results from this are very stimulative and prove that the system could be used in a practice in the near future.

The structure of this paper is as follows. In Section 2, the problem of drafting in Ironmans is discussed. Section 3 describes the proposed concept of drafting detection system. In Section 4, experiments and results of using the test system for drafting detection are presented. The paper concludes with a discussion about the performed work and outlining directions for further development.

%------------------------------------------------------------------------------------------
\section{Drafting violation in Ironmans}
Ironman remains to be the most prominent triathlon among its different kinds. It first appeared in 1978 when a group of enthusiastic athletes tried to perform three marathons (swimming, cycling and running) in one day. In order to honor this event, a World Championship in Ironman is organized each year in Hawai. Ironman is the hardest one-day trial~\cite{rauter:2011} in the world attracting more and more participants each day.

\begin{figure*}[htb]
%\vspace{-5mm}
    \begin{center}
        \includegraphics[scale=0.7]{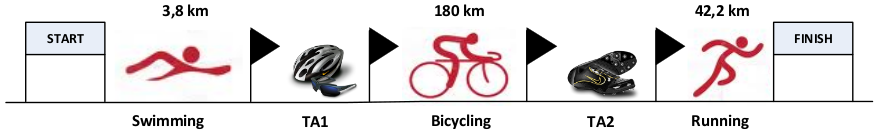}      %
        \caption{Triathlon Ironman}
        \label{pic:IM}
    \end{center}
\vspace{-5mm}
\end{figure*}

A classical Ironman is illustrated in Fig.~\ref{pic:IM} showing that the competition is performed sequentially one discipline after another. After a common start, the competitors start with swimming, then ride their bicycles, and finish with running. There are two transition areas, i.e. TA1 and TA2 (Fig.~\ref{pic:IM}) in the course: competitors first discard their swimming wear and prepare themselves for cycling and then they leave their bicycles and then prepare themselves for running. The times the competitors spend in the transition areas last less than two minutes. However, these times are added to their results from all the three marathon disciplines to assessing their finishing achievements.

Drafting violation usually appears during cycling and denotes a phenomenon by which one competitor purposely rides bicycle directly behind the other, and thereby avoids wind resistance. The drafting competitor can speed-up the average riding velocity and at the same time saves his energy. Usually, most competitors ride their bicycles together in groups. Such riding does not reflect their real efficiency and is regulated by the WTC using the following rules~\cite{wtc:2010}:
\begin{itemize}
  \item Drafting of other competitors or vehicles is forbidden.
  \item A competitor must ride his bicycle over a distance of almost 7 meters, except in case of overtaking.
  \item Overtaking occurs when the front wheel of the overtaking competitor overtakes the front wheel of the competitor being overtaken.
  \item A competitor must be overtaking on the left side for no more than 20 seconds. After this time the overtaking competitor must return to the right side of the road.
  \item The overtaken competitor must fall back 7 meters behind the overtaking competitor before he can start to overtake a competitor in front of him.
\end{itemize}

In drafting violation shown in Fig.~\ref{pic:IM2} competitor B rides his bicycle directly behind competitor A at a distance of 5 meters. If competitor B does not increase this distance to 7 meters in 20 seconds, the referee can punish him for drafting violation. As seen from Fig.~\ref{pic:IM2}, the drafting zone behind competitor A is determined by a rectangle $2 \times 7$ meters, i.e. one meter left and one right of competitor A. Therefore, if competitor B stays for more than 20 seconds inside this virtual rectangle, his riding will be announced as aa act of drafting violation.

\begin{figure}[htb]
%\vspace{-5mm}
    \begin{center}
        \includegraphics[scale=0.55]{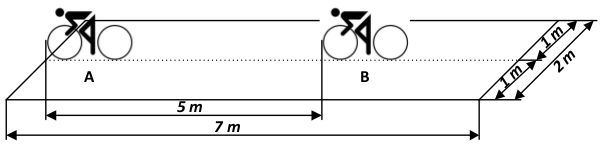}      %
        \caption{Drafting violation}
        \label{pic:IM1}
    \end{center}
\vspace{-5mm}
\end{figure}

The phenomenon of drafting violation is detected and punished by referees during official Ironmans. However, because of too many competitors on the cycling course, the referees have several problems, for example:
\begin{itemize}
  \item they can only deal with a particular section of the bicycle course at a time,
  \item distances between competitors are estimated by their senses, and
  \item the duration of drafting violation is left to their subjective judgements.
\end{itemize}

Pervasive computing can be very helpful when looking for an automatic solution. That is, it allows for dealing with all competitors on the bicycle course simultaneously. On the other hand, a GPS receiver enables precise determination of the distances between competitors and duration times of any drafting violation. Note that an attempt has been made to include the mentioned advantages of pervasive computing into this concept of drafting detection system in Ironmans.

\section{Concept of drafting detection system in Ironmans}
\begin{figure}[htb]
%\vspace{-5mm}
    \begin{center}
        \includegraphics[scale=0.44]{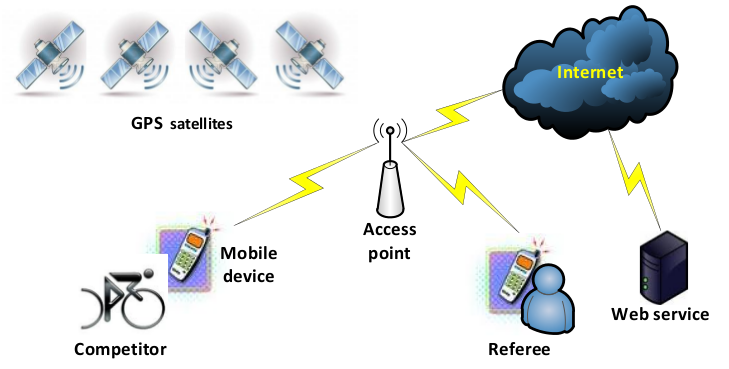}      %
        \caption{Concept of drafting detection in Ironmans}
        \label{pic:IM2}
    \end{center}
\vspace{-5mm}
\end{figure}

Our concept of drafting detection system in Ironmans is presented in Fig.~\ref{pic:IM2}. It is shown that the mobile device represents the basis of the system. It is borne by the competitor on his bicycle. It provides information about the competitor's current position and transmits it over a wireless modem to a web service. The web service assesses whether a competitor is violating the drafting rules. Furthermore, violations can be tracked by the referees on motorcycles over similar mobile devices.

The proposed concept consists of four elements:
\begin{itemize}
  \item GPS receiver,
  \item wireless modem,
  \item pervasive client of web services, and
  \item web service.
\end{itemize}
Note that the today's smart-phones are already equipped with GPS receivers and wireless modems. There are operating system also running on these devices, for example Windows Mobile, BlackBerry, MacOS, Android, etc. Also, there is a platform enabling development of the web service in higher-level programming languages installed in smart-phones. The Android platform~\cite{brunette:2010,darcey:2011} used in our work, for example, allows for development of pervasive applications~\cite{fling:2009} the programming language Java~\cite{bell:2010} and framework Eclipse~\cite{abbott:2009}. When calculating the distances between competitors, the precision is affected by the precision of the GPS receiver. This is why in the remainder of our paper navigation technologies will be presented in detail. This section concludes with a description of the web service for drafting detection.

\subsection{GPS, DGPS and Galileo navigation systems}
GPS is a global positioning system developed in 1973 by the U.S. Department of Defence~\cite{dierendonck:1996} and based on a set of broadcasting satellites orbiting around the Earth. A GPS receiver uses these satellites as its reference points for calculating positions of objects on the Earth. A connection with almost four broadcasting GPS satellites is necessary for the calculations. Each satellite broadcasts messages containing information about its precise position. The GPS receiver uses these messages for calculating its position on the Earth using a triangulation method~\cite{zalik:2005}. This position is usually represented as geographical coordinates, i.e. longitude, latitude and altitude.

GPS consists of three segments: a space, user, and control. The space segment consists of 24 to 31 satellites orbiting in a GPS constellation with a rotation of the Earth at an altitude of around 20 kilometers. The user segment consists of GPS receivers (for example, Garmin, Polar, Sunto, smart-phones, etc.). The control segment ensures that the GPS satellites work correctly and efficiently.

GPS enables a two-level service~\cite{misra:2010}:
\begin{itemize}
  \item standard positioning (SPS) and
  \item precise positioning (PPS).
\end{itemize}

The SPS system positions objects on the Earth with a precision of 20 meters and is intended for common usage. A PPS system is more precise. It can position objects with a precision of only a few centimeters, but is only intended for a military use.

The SPS system cannot be employed for the precise positioning of objects on the Earth because of various reasons that might impact the GPS performance~\cite{agnew:2007}. These reasons include effects of the ionosphere and troposphere, unscheduled satellite failures, satellite unavailability due to the scheduled maintenance, repairs, repositioning, and testing~\cite{prasad:2005}. These factors could result in errors caused by broadcasting the GPS positions of satellites. As a result, these errors might negate affect accuracy, availability, and security of GPS information. Therefore, a differential GPS (DGPS) has emerged~\cite{prasad:2005} based on correction signals broadcasted by transmitters on geostationary satellites. This method clearly increases the precision of the SPS system (up to 5 meters). Note that in Europe the EGNOS system is employed, whilst in the U.S. it is the WAAS system.

Galileo is a global navigation system~\cite{prasad:2005} currently under construction. It is being developed by the protection of the European Union, i.e. European Space Agency (ESA). The purpose of the Galileo project, valued at 20 milliard EURos, is to create a navigation system that would ensure Europe high precision positioning and would be independent from the American GPS or Russian GLONASS systems. Galileo is planned to begin its operating in 2012. It will have two control centers, i.e. in Germany and Italy. Moreover, it will consist of 30 satellites orbiting at an altitude of 23 kilometers. During free-of-charge usage, it will ensure positioning of up to one meter precisely, whilst for commercial use, it will be more precise, i.e. up to one centimeter.

\subsection{Web service for drafting detection}
The simplest way to connect a pervasive client to the web server is over web services. The System Oriented Architecture (SOA)~\cite{alonso:2010} is a de-facto standard for message exchanging between web services. SOA consists of four protocols:
\begin{itemize}
  \item Extensible Markup Language (XML): intended for data tagging,
  \item Service Oriented Architecture Protocol (SOAP): is a message transfer protocol,
  \item Web Service Description Language (WDSL): used for describing available services on the Internet,
  \item Universal Description, Discovery and Integration (UDDI): searching for web services on the Internet.
\end{itemize}

The development of web services is difficult because of several protocols the programmer must be aware of. Therefore, Apache developed a tool AXIS2~\cite{jayasinghe:2011} that allows the programmer to work with web services on a higher level. This tool, which is also integrated into Eclipse framework, was used in the development of our system concept.

Through there are several architectures available today to connect mobile devices and SOA, direct calling the web service was employed in our concept, i.e. all demands from the mobile device are sent to the web service directly. That is, all the packing and unpacking of SOAP messages are performed by a pervasive client alone. However, these functionalities are enabled by Android automatically by means of the KSOAP2 library.

Information about the GPS positions of competitors sent from the pervasive client on a mobile device are written in geographical coordinates as tuples $\langle lon, lat, alt \rangle$, where $lon$ denotes the longitude, $lat$ the latitude, and $alt$ the altitude. With geographical coordinates they are, however, hard to calculate. Therefore, they need to be transformed into a three-dimensional metric coordinate system. In geography, such a coordinate system is the Universal Traverse System (UTM)~\cite{navy:1989}. The UTM coordinate system represents a Mercator projection of the Earth to a plane divided into 60 latitudinal and 30 longitudinal zones. Each position in this coordinate system is represented as a quadruple $\langle lat\_zone, lon\_zone, east, north \rangle $, where $lat\_zone$ and $lon\_zone$ denote the number of latitudinal and longitudinal zones, whilst $east$ is the projected distance from the central meridian, and $north$ the projected distance from the equator. Note that both values $east$ and $north$ are defined in meters. Although the basis of this transformation represents the elementary trigonometric and algebraic functions, transformation equations are actually very complex~\cite{vincenty:1975, teunissen:1999}. Therefore, in this work, it was decided to employ transformation of the author Salkosua~\cite{sami:2007}, implemented in Java. The Euclidian distance was used to calculate the distances between competitors. The Euclidian distance in a 2-dimensional space was employed because the positions of competitors were captured at one second intervals.

The algorithm used in drafting detection is relatively easy. Each record representing the competitor's current GPS position can be defined as tuple $\langle i,x,y,z,t,l \rangle$, where $i$ denotes the competitor's starting number, $x,y,z$ his current position within the coordinate system UTM, $t$ the registration time of the record, and $l$ the calculated path length. Calculated path length $l$ is obtained by projecting the current position of the $i$-th competitor on the line that connects the points gained by tracking the cycling course with a precise GPS device, at each second. This course tracking must be performed before starting the bicycle race.

Path length $l$ impacts the current placing of a competitor. That is, the longer the path length, the better the competitor's placing. However, this placing is determined by sorting competitors with regard to decreasing path length $l$. Competitors can be placed in neighborhood of the $i$-th competitor when capturing their GPS positions is dense. As a result, the number of exchanges needed for sorting is small.

After placing the $i$-th competitor, it is necessary to check how far he is from the nearest competitor behind him (the (i+1)-th competitor). If the distance between the $i$-th and the $(i+1)$-th competitors is less than 7 meters for more than 20 seconds, the drafting detection system announces drafting violation by the $(i+1)$-th competitor.

\section{Experiments and Results}
The goal of our experimental work was to show that the today's mobile devices can be used in drafting detection in Ironmans. Two experiments were conducted to prove our idea:
\begin{itemize}
  \item comparing the reference distances on the Earth with the distances calculated using information from GPS receivers, and
  \item simulating drafting detection in practice.
\end{itemize}

14 co-linear points on the Earth were selected for the first experiment. The distance between the two nearest points was one meter. Then, a walk with various mobile devices through the reference points was made and their GPS positions measured. The distance to the starting point was calculated from the GPS positions. Note that this walking was done at a uniform speed of 1 cm/sec. In this test we were interested in establishing how sensitive mobile devices are at the distances of 7 meters where drafting violation can occur. The following devices were used in our experiment:
\begin{itemize}
  \item Samsung Galaxy smart-phone,
  \item HTC Wildfire smart-phone,
  \item U-blox device connected to a mobile PC with USB interface, and
  \item Garmin Etrex-H device.
\end{itemize}

All the used devices capture GPS positions at one second intervals. Garmin Etrex-H logs these positions into an internal storage that can be copied onto a PC in the form of .GPS files. Similarly, the U-blox device archives the GPS position using the same file format. In contrast, information from smart-phones can be obtained online.

\begin{figure}[htb]
%\vspace{-5mm}
    \begin{center}
        \includegraphics[scale=0.6]{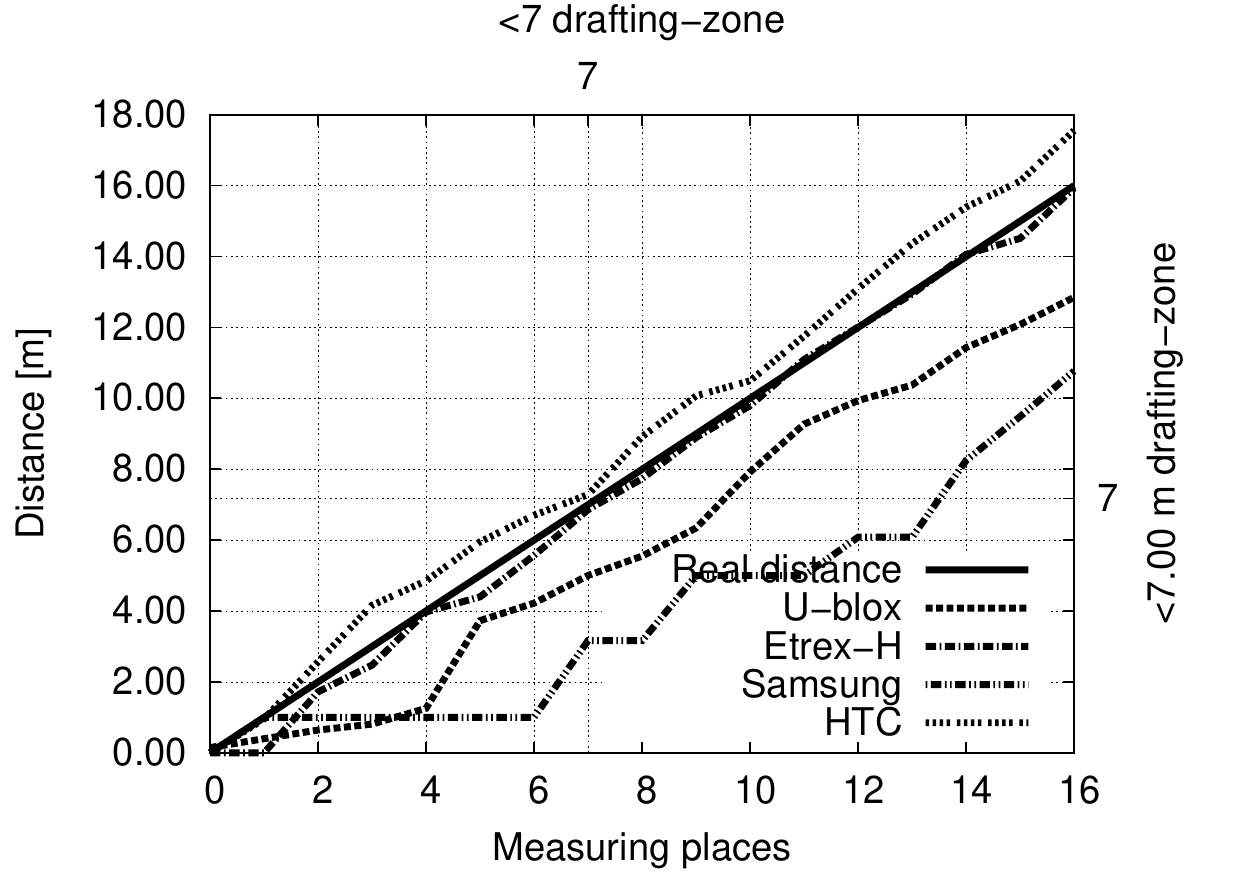}      %
        \caption{Comparing the reference distances with the calculated ones}
        \label{pic:test}
    \end{center}
\vspace{-5mm}
\end{figure}

Results of our experiments are presented in Fig.~\ref{pic:test} in which the raw line in graph connects the real distances of the reference points on the Earth, whilst the lines denoted by labels $" < 7 "$ define the drafting zone. It can be seen that the Garmin Etrex-H device is the nearest to the real distances. As the HTC Wildfire device provides excellent results compared to other smart-phones, it could be the best candidate for being uses in practice.

\begin{figure}[htb]
%\vspace{-5mm}
    \begin{center}
        \includegraphics[scale=0.5]{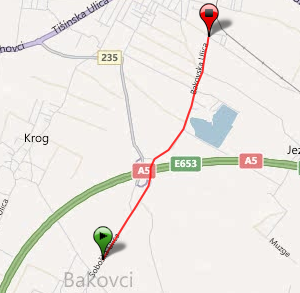}      %
        \caption{Course of the bicycle race (Powered by Google Map)}
        \label{pic:trasa}
    \end{center}
\vspace{-5mm}
\end{figure}

In our second experiment we simulated a drafting violation act. Two competitors ere racing on a 3.332 kilometer long bicycle course (Fig.~\ref{pic:trasa}). The course was relatively easy with no slopes. Each competitor was equipped with the Garmin Forerunner 110 sport watches that have very precise DGPS receivers. Information about the riding was copied from sport watches in .GPS formatted files.

An agent was developed to read the files and transfer information in real-time to the drafting detection system. As smart-phones are awkward when riding and need connection with a public wireless network, which is expensive, sport watches were used in our experiment.

\begin{figure}[htb]
%\vspace{-5mm}
    \begin{center}
        \includegraphics[scale=0.7]{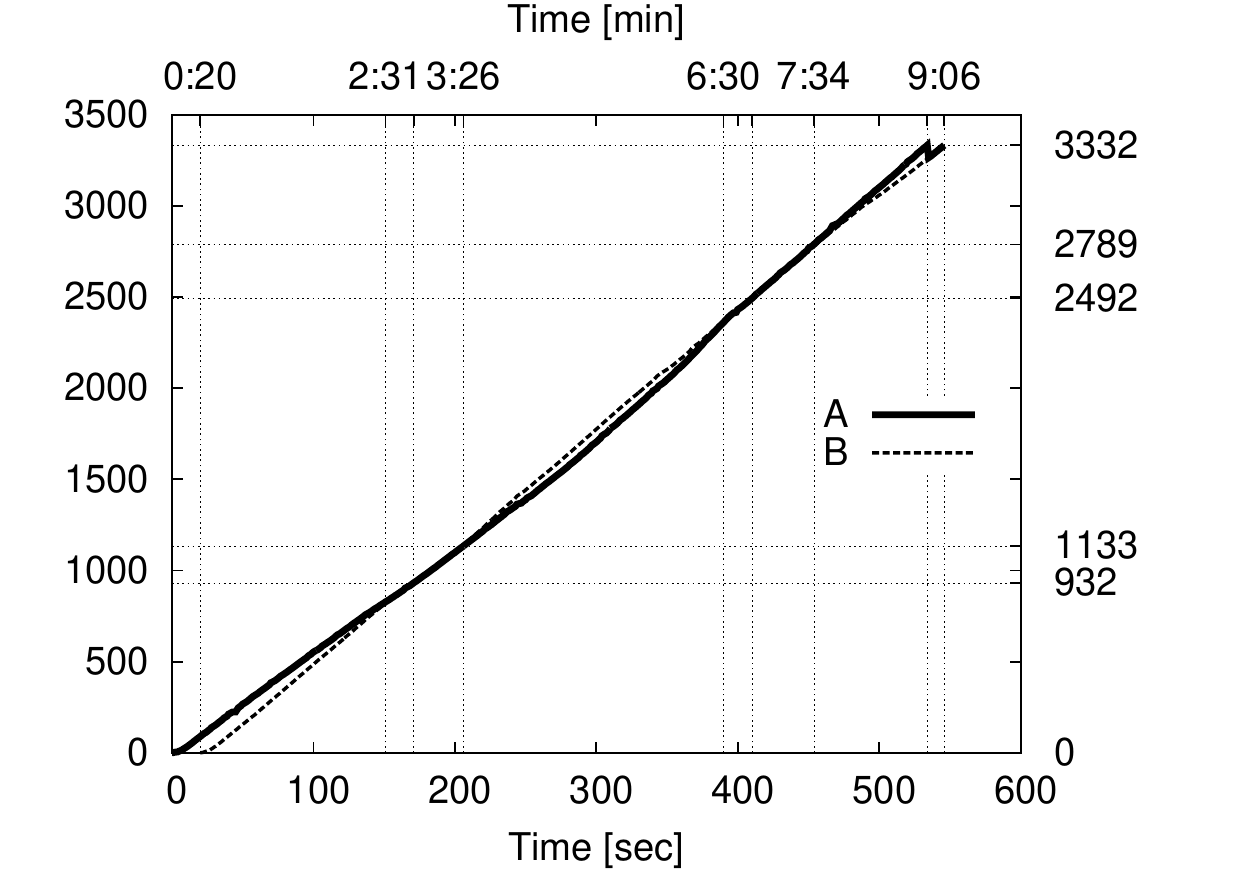}      %
        \caption{Simulation of drafting violation}
        \label{pic:dirka}
    \end{center}
\vspace{-5mm}
\end{figure}

As seen from racing of competitors A and B in Fig.~\ref{pic:dirka}, competitor A started 20 seconds before competitor B. After 2:31 minutes, competitor B was riding behind competitor A and, after further 20 seconds, he started to violate the drafting-rules. His violation was correctly detected by the drafting detection system. After 200 meters of drafting violation, competitor B overtook competitor A and rode in front of him at a distance of 20 to 30 meters. The situation was changed after 6:30 minutes. That is, competitor A rode into the drafting zone of competitor B for almost 300 meters, then, he overtook him and finished the race in a time of 8:54 minutes, i.e. 12 seconds before competitor B. The second competitor was 8 seconds quicker than the first one because competitor A started 20 seconds before competitor B.

In general, it can be concluded that drafting violations can be detected by using the today's mobile devices equipped with GPS receivers. Though smart-phones offers these functionalities, they are unsuitable for being used in cycling races. A special problem is their high usage of power that, in practice, does not even cover the need of the whole bicycle race (e.g., 6 hours at the speed of 30 km/h). So, mobile devices with built-in precise GPS receivers and efficient wireless modems are believed to be more suitable.

\section{Conclusion}
The paper shows that drafting detection in Ironmans is not an illusion and likely to be used in the near future. This conclusion is based on several facts. For example, the Galileo navigation system is already completed and will allow positioning of objects on the Earth at a precision rate of up to one centimeter. The appearance of 4G mobile networks~\cite{prasad:2010} has increased efficiency of mobile devices. The rapid growth of the Internet demands more efficient web servers. Pervasive computing has also grown very fast. Several providers have emerged on the market offering specialized pervasive devices combining the precise GPS receivers with wireless modems into mobile devices that are lightweight, inexpensive, and of small power consumption. This kind of mobile devices are believed to be a solution to the above discussion.

Our further work will be towards integrating the drafting detection system with a timing system controlled by the EasyTime~\cite{fister:2011,fister:2011a,fister:2011b} domain-specific language.

\bigskip{\small \smallskip\noindent Updated 8 June 2012.}
\end{document}